\documentclass[%
 reprint,
superscriptaddress,
 amsmath,amssymb,
pra,
]{revtex4-2}

\usepackage{graphicx}
\usepackage{dcolumn}
\usepackage{bm}
\usepackage{braket}
\usepackage{titlesec}
\usepackage{relsize}
\begin{document}

\preprint{APS/123-QED}

\title{Realization of Backward Retrieval in a Stark-modulated Spin-wave Quantum Memory}

\author{Zhenqi Xu}
\author{Mucheng Guo}
\author{Weiye Sun}
\affiliation{Southern University of Science and Technology, Shenzhen 518055, China.}
\affiliation{International Quantum Academy, and Shenzhen Branch, Hefei National Laboratory, Shenzhen, 518048, China}

\author{Pengjun Liang}
\author{Zongquan Zhou}
\affiliation{CAS Key Laboratory of Quantum Information, University of Science and Technology of China, Hefei 230026, China}
\affiliation{Anhui Province Key Laboratory of Quantum Network,  University of Science and Technology of China, Hefei 230026, China}
\affiliation{CAS Center for Excellence in Quantum Information and Quantum Physics, University of Science and Technology of China, Hefei 230026, China}
\affiliation{Hefei National Laboratory, University of Science and Technology of China, Hefei 230088, China}

\author{Fudong Wang}
\email{fdwang.phys@foxmail.com}
\author{Shuping Liu}
\email{liushuping@iqasz.cn}
\author{Manjin Zhong}
\email{manjin.zhong@gmail.com}
\affiliation{International Quantum Academy, and Shenzhen Branch, Hefei National Laboratory, Shenzhen, 518048, China}

\begin{abstract}

We report the first experimental realization of backward retrieval in a spin-wave quantum memory based on a Stark-echo-modulated protocol in Eu$^{3+}$:Y$_2$SiO$_5$. By using Stark control, we preserve the full optical depth of the ensemble while suppressing coherent noise, enabling conditional storage fidelities above 97\%. Our analysis shows that the present backward-retrieval efficiency is mainly limited by technical imperfections rather than by fundamental constraints. With realistic engineering improvements, backward retrieval in this protocol could move beyond the reabsorption-limited forward-emission regime. The protocol is also compatible with cavity-enhanced operation, offering an additional route toward higher efficiencies. These findings establish Stark-echo modulation as a practical and scalable route to high-efficiency, long-lived solid-state quantum memories.

\end{abstract}

\maketitle

\section{Introduction}

Quantum networks aim to distribute quantum states among distant nodes, forming the basis for secure communication, high-precision sensing and distributed quantum computation~\cite{kimble2008quantum,hammerer2010quantum,Knill2001A,lei2025deterministic}. A central component of such networks is the quantum memory, which synchronizes probabilistic photon events by mapping flying qubits onto long-lived matter states and retrieving them on demand~\cite{heshami2016quantum,simon2010quantum,briegel1998quantum,duan2001long,Rakonjac2021Entanglement}. Rare-earth-ion (REI)-doped crystals combine long optical coherence at cryogenic temperatures with spin states that can maintain coherence for orders of magnitude longer. This combination provides a powerful platform for efficient absorption and reemission of photonic excitations, as well as for long-lived, on-demand quantum storage \cite{moiseev2003quantum, nilsson2005solid, tittel2010photon, zhong2015optically, wang2025nuclear}. Their spectral multiplexing capability \cite{Lago2023long,tang2015storage}, and compatibility with photonic integration \cite{Rakonjac2022storage,Liu2020on} further enhance their suitability for high-performance quantum memories.

A variety of quantum-memory protocols have been developed to harness the long-lived spin states of REI-doped crystals~\cite{moiseev2001complete,moiseev2004possibilities,moiseev2003quantum,ma2021one,pierre2015Coherent,Timoney2012Atomic}. Exploiting these spin states requires transferring optical excitations to spin-waves while maintaining sufficient optical depth to ensure high storage and retrieval efficiency \cite{gundougan2015solid,Timoney2012Atomic,alqedra2024stark}. However, reabsorption, stemming from the directionality of the emitted echo, fundamentally limits the theoretical storage efficiency to 54\%. Backward retrieval avoids reabsorption and can, in principle, reach near-unity efficiency~\cite{sangouard2007analysis}. Leading schemes such as spin-wave atomic frequency comb (AFC)~\cite{afzelius2009multimode,jobez2015coherent} and noiseless photon-echo (NLPE)~\cite{liu2025millisecond,zhu2024integrated,ma2021elimination} either reduce effective optical depth during spectral tailoring or impose geometric constraints that hinder true backward emission. Consequently, a practical quantum-memory protocol that  preserves full optical depth and enables low-noise backward retrieval remains elusive.

In this work, we experimentally demonstrate a Stark-echo (SE)-modulated spin-wave quantum memory in Eu${^{3+}}$:Y$_2$SiO$_5$, with both forward and backward retrieval~\cite{inprepare2}. This protocol combines the noiseless photon-echo (NLPE) scheme~\cite{ma2021elimination,beavan2011photon} with Stark-control techniques~\cite{arcangeli2016stark,macfarlane2014optical}, allowing deterministic scrambling and rephasing of the ensemble phases without sacrificing optical depth~\cite{arcangeli2016stark}. As a result, it suppresses coherent noise and provides the geometric flexibility needed to realize backward emission. We measure conditional storage fidelities exceeding 97\% for both retrieval directions, identify key factors limiting backward efficiency, and outline clear routes for substantial improvements. These results establish Stark-echo modulation as a practical foundation for high-efficiency, scalable REI-based quantum memories suitable for future quantum-network applications ~\cite{Zhu2020Coherent,liu2025millisecond}.

\begin{figure*}
	\includegraphics[width=\linewidth]{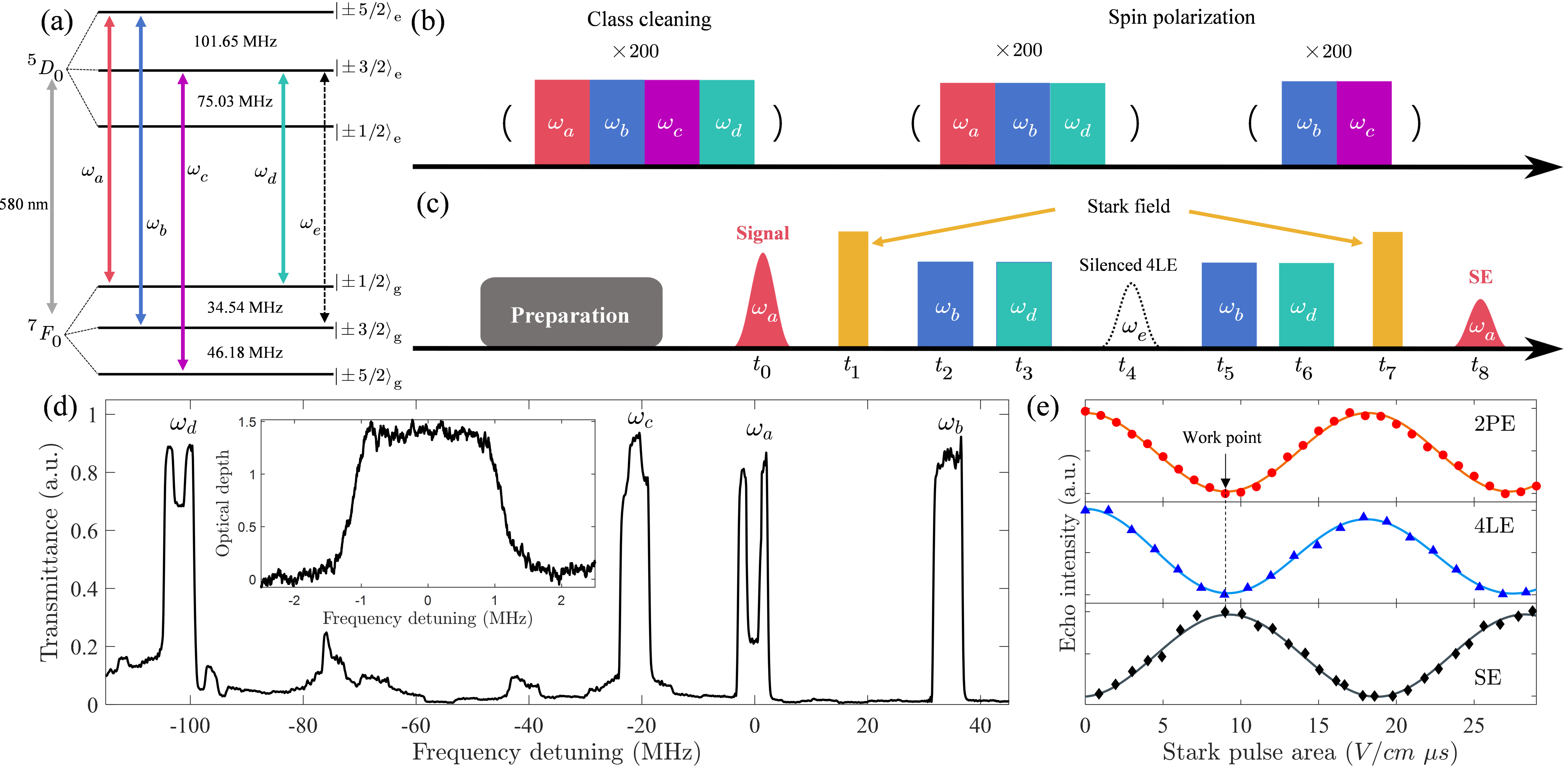}
	\caption{\label{fig:fig1}(a) Energy level structure of ${}^{151}\mathrm{Eu}^{3+}$ used in $\mathrm{Eu}^{3+}:\mathrm{Y}_2\mathrm{SiO}_5$ crystal. (b) Preparation sequence including class cleaning and spin polarization. (c) Full quantum memory sequence showing the preparation process, four optical $\pi$-pulses and two electric field pulses. (d) Optical spectrum after class cleaning and spin polarization; inset: the resulting absorption feature. (e) Stark modulation measurements via two-pulse photon echo (2PE), four-level photon echo (4LE), and the proposed Stark echo (SE) method, respectively. Arrows and dashed lines highlights the electric field pulse area of $9.25~\mathrm{V/cm \cdot \mu s}$ used for the memory operation. Data points indicate experimental values, solid curves are fitting results.}
\end{figure*}

\section{Theory}

The SE protocol exploits the inversion symmetry of the ensemble together with the linear Stark effect. In certain low-symmetry solid-state systems, inversion-related sub-ensembles can experience linear Stark shifts of opposite signs under an applied electric field~\cite{Kaplyanskii2002Linear,macfarlane2007optical}. In Eu$^{3+}$:Y$_2$SiO$_5$, this gives rise to two classes of $\mathrm{Eu}^{3+}$ ions whose optical transition frequencies shift in opposite directions:
\begin{equation}
	\Delta \nu = - L\,\Delta\boldsymbol{\mu}\cdot\mathbf{E},
\end{equation}
where $\Delta\boldsymbol{\mu}$ denotes the difference in permanent electric dipole moments between the two levels, and $L$ is the local-field correction factor~\cite{mahan1967local}. Under a Stark pulse of duration $\tau_S$ and electric field $\mathbf{E}_s$, the two inversion-related sub-ensembles accumulate opposite phases, leading to a relative phase difference~\cite{macfarlane2014optical}
\begin{equation}
	2\phi_S = 2\frac{\Delta\boldsymbol{\mu} \cdot \mathbf{E}_s}{\hbar}\,\tau_S.
\end{equation}
This relative phase controls the collective emission of the ensemble: when $2\phi_S=\pi$, the two sub-ensembles interfere destructively and the echo is silenced, whereas when $2\phi_S=2\pi$, they rephase and the collective emission is restored.

The SE protocol requires a four-level scheme, which is implemented here using the Eu$^{3+}$ energy-level structure in Eu$^{3+}$:Y$_2$SiO$_5$, as shown in Fig.~\ref{fig:fig1}(a). Figure~\ref{fig:fig1}(b) shows the preparation sequence used to initialize the ions in $\ket{\pm 1/2}_g$, thereby enabling the implementation of the protocol. The pulse sequence shown in Fig.~\ref{fig:fig1}(c) consists of four optical $\pi$ pulses interleaved with two Stark pulses. The optical coherence created by the input signal is transferred by the control $\pi$ pulses into spin coherence. During the interval from $t_2$ to $t_3$, the coherence is stored on the ground-state spin transition $\ket{\pm 1/2}_{\mathrm{g}} \leftrightarrow \ket{\pm 3/2}_{\mathrm{g}}$, and from $t_5$ to $t_6$ it is transferred to the excited-state spin transition $\ket{\pm 3/2}_{\mathrm{e}} \leftrightarrow \ket{\pm 5/2}_{\mathrm{e}}$. The ground-state spin coherence thus provides the degree of freedom for long-lived storage. A Stark pulse applied at $t_1$ induces a phase difference of $2\phi_S=\pi$ between the two classes of ions, thereby scrambling the ensemble coherence and silencing the unwanted 4LE emission~\cite{beavan2011photon}. A second Stark pulse at $t_7$ rephases the accumulated shifts to $2\phi_S=2\pi$, producing an SE at $t_8$ whose emission direction is determined by the phase-matching condition~\cite{ma2021elimination}:
\begin{equation}\label{eqn:pm}
	\mathbf{k}_{\text{echo}} = \mathbf{k}_0 - \mathbf{k}_2 - \mathbf{k}_3 + \mathbf{k}_5 + \mathbf{k}_6,
\end{equation}
where $\mathbf{k}_i$ denotes the wave vector of the optical pulse applied at time $t_i$. The emission time of the retrieved echo is given by $t_8 = t_0 - t_2 - t_3 + t_5 + t_6$.

\begin{figure}[t]
	\includegraphics[width=\linewidth]{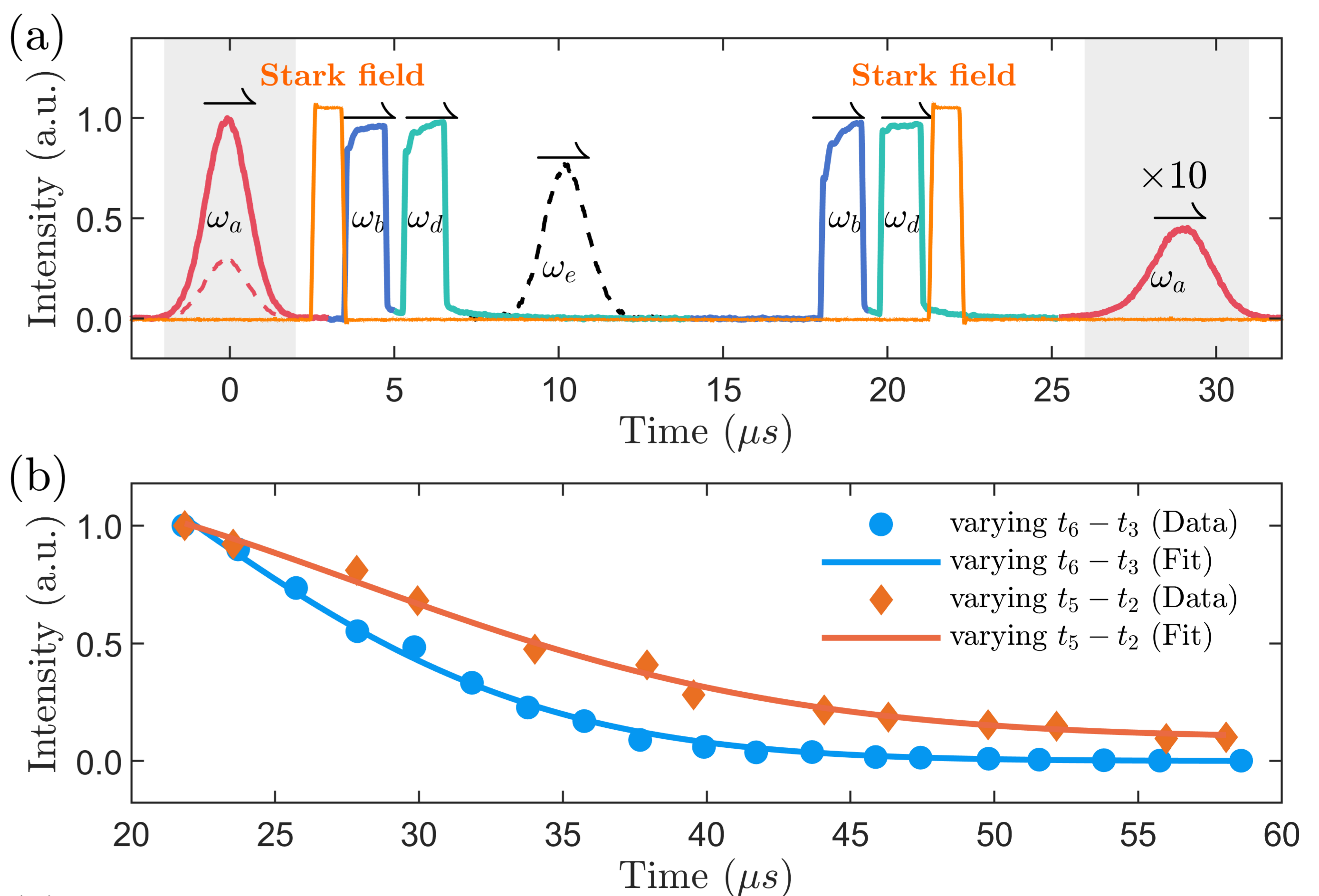}
	\caption{\label{fig:fig2}Forward retrieval. (a) Experimental sequence with applied propagation direction (shown as arrows) of each pulse. The black dashed curve represents the suppressed 4LE. The output signal in the shaded area is magnified by a factor of 10 for clarity. (b) SE decay curves measured by varying $t_6-t_3$ (blue) and $t_5-t_2$ (orange). }
\end{figure}

By adjusting the propagation direction of the $\pi$-pulse at $t_5$ or $t_6$, the retrieved echo can be directed either forward or backward. Neglecting the inhomogeneity of the Stark pulses, we adopt an effective model based on Ref.~\cite{ma2021elimination} to describe the experimental data. In this simplified treatment, finite-bandwidth, spectral-dispersion, and propagation effects are neglected~\cite{moiseev2012rephasing,arslanov2017optimal,moiseev2025optical}, and the total memory efficiency is given by
\begin{equation}
	\eta_{_\text{\tiny SE}} = \eta_{\text{\tiny retrieval}} \, \eta_{\text{\tiny pm}} \,
	 \eta_{\text{\tiny control}}^4\,
	\eta_{\text{\tiny decay}},
	\label{eq:eff}
\end{equation}
where $\eta_{\text{\tiny retrieval}}$ is the retrieval efficiency, taking the form $\eta_{\text{\tiny retrieval}} = d^2 e^{-d}$ for forward retrieval and $\eta_{\text{\tiny retrieval}} = (1 - e^{-d})^2$ for backward retrieval, with $d=\alpha L$ the optical depth. The factors $\eta_{\text{\tiny pm}}$, $\eta_{\text{\tiny control}}$, and $\eta_{\text{\tiny decay}}$ represent the phase-matching efficiency, the population-transfer efficiency of each $\pi$-pulse,  and the decoherence term, respectively. Decoherence includes both optical and hyperfine contributions and is used here as a practical fitting model for the measured decay curves:
\begin{equation}
	\eta_{\text{\tiny decay}} =  e^{ -\frac{\Gamma_{13}^2 (t_5 - t_2)^2}{2 \ln 2 / \pi^2} } \cdot  e^{ -\frac{\Gamma_{35}^2 (t_6 - t_3)^2}{2 \ln 2 / \pi^2} - 2 \gamma (t_6 - t_3) },
	\label{eq:coherence}
\end{equation}
where $\Gamma_{13}$ and $\Gamma_{35}$ are the inhomogeneous broadening of the spin transitions $\ket{\pm 1/2}_\mathrm{g} \leftrightarrow \ket{\pm 3/2}_\mathrm{g}$ and $\ket{\pm 3/2}_\mathrm{e} \leftrightarrow \ket{\pm 5/2}_\mathrm{e}$, respectively, and $\gamma$ is the effective optical decoherence rate.  

\section{Experiment}

\begin{figure}[t]
	\includegraphics[width=1.0\linewidth]{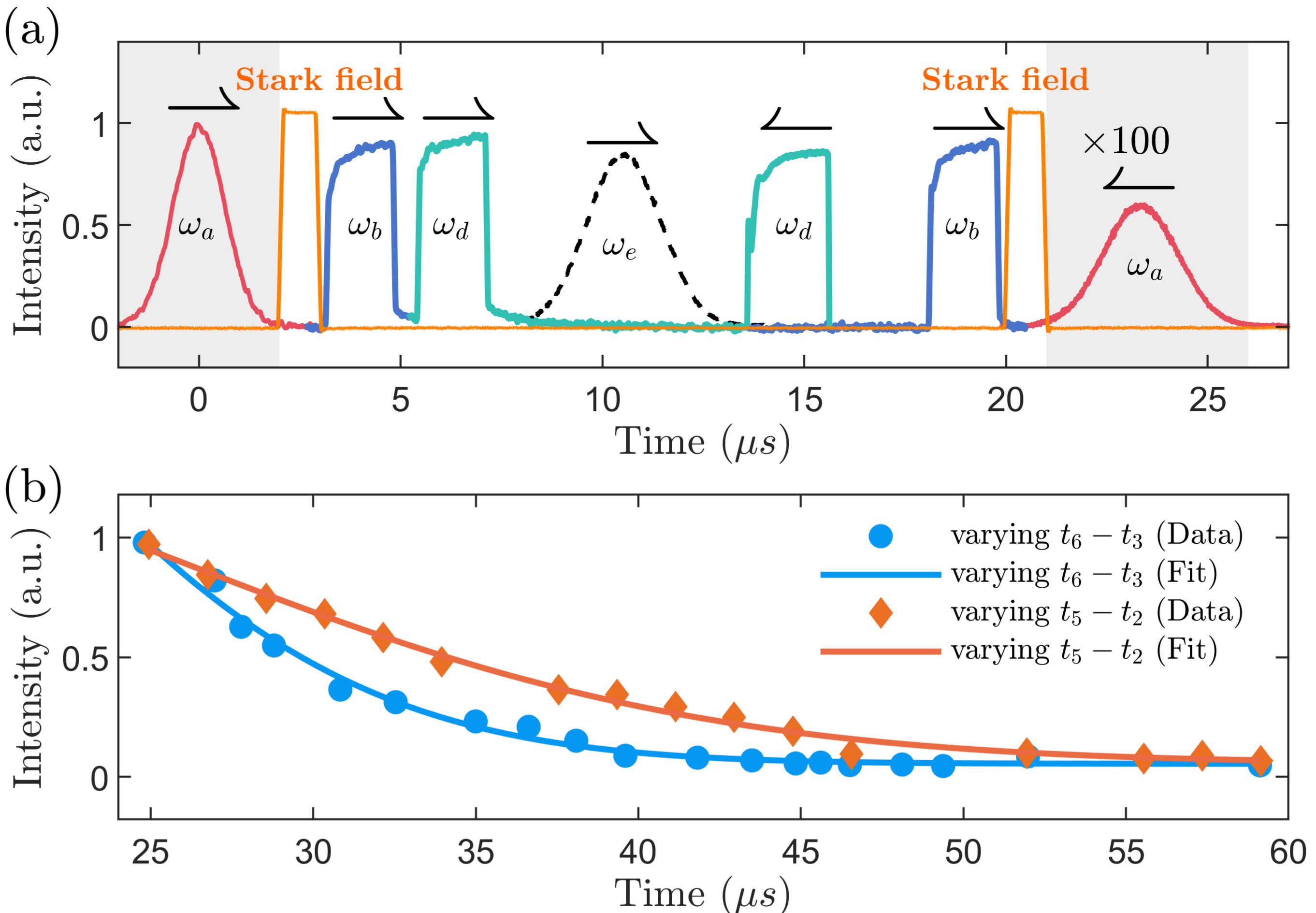}
	\caption{\label{fig:fig3} Backward retrieval. (a) Experimental sequence with the propagation directions (shown as arrows) of different pulses and the silenced 4LE noise showing by the black dashed line. The output signal in the shaded area is magnified by a factor of 100 for clarity. (b) SE decay profiles obtained by varying $t_6-t_3$ (blue) and $t_5-t_2$ (orange). }
\end{figure}

The experiment was performed on a 0.1\% Eu$^{3+}$:Y$_2$SiO$_5$ crystal with dimensions of 5 mm $\times$ 5 mm $\times$ 10 mm along the $D_1$, $D_2$, and $C_2$ axes. We used the $\mathrm{Eu^{3+}}$ ions located at site 1 of the crystal and the relevant hyperfine structure is shown in Fig.~\ref{fig:fig1}(a). All the optical transitions were excited with light propagating along the $C_2$ axis with  the polarization along $D_1$ axis. The electric fields were applied parallel to $D_1$ axis via copper electrodes attached to the crystals using silver paint. The crystal was cooled to 1.8 K in a cryostat. The experiment employed a collinear counter-propagating configuration, in which a fiber coupler was used to facilitate alignment. An acousto-optic modulator (AOM) was used to gate the detection window of the avalanche photodiode (APD), while two separate AOMs were used to generate and modulate the optical pulses. More details of the experimental setup are presented in Supplemental Materials~\cite{Sup}.

The preparation sequence, including class cleaning and spin polarization~\cite{lauritzen2012spectroscopic}, is illustrated in Fig.~\ref{fig:fig1}(b). Using 1-ms frequency-chirped pulses with a peak power of 100~$\mu\mathrm{W}$, this sequence produced the absorption feature shown in the inset of Fig.~\ref{fig:fig1}(d), with an optical depth of 1.3 and a bandwidth of 2~MHz. The linear Stark shift was then determined using two-pulse echoes (2PE)~\cite{macfarlane2014optical} and four-level echo (4LE) sequences~\cite{beavan2011photon}, as shown in Fig.~\ref{fig:fig1}(e). The measured Stark coefficient, $27.5 \pm 0.2$~kHz/(V$\cdot$cm$^{-1}$), agrees well with previous reports~\cite{macfarlane2014optical,zhang2020precision}. SE rephasing was further verified by applying a fixed Stark pulse at $t_1$ and varying the area of the second pulse at $t_7$ [bottom panel of Fig.~\ref{fig:fig1}(e)].

\section{Results and Discussion}

\subsection{Echo retrieval}

To verify the operation of the SE protocol, forward retrieval was first experimentally demonstrated by direct detection of the SE intensity with an avalanche photodiode (APD), as shown in Fig.~\ref{fig:fig2}(a). At early times, the SE was primarily obscured by the 4LE background from the input signal and $\pi$ pulses, while the 2PE noise induced by the control pulses was suppressed by the electric field. By adjusting the relative timing of the $\pi$ pulses, we separated the SE from the background and measured a memory efficiency of $4.8 \pm 0.06\%$ at $29~\mu\mathrm{s}$. The relevant noise processes are analyzed in the next subsection.

Coherence dynamics were characterized by recording the SE decay curves via heterodyne detection, as shown in Fig.~\ref{fig:fig2}(b). Fitting the data using Eq.~(\ref{eq:eff}) and Eq.~(\ref{eq:coherence}) yielded linewidths of $\Gamma_{35} = 21.9 \pm 1.0 ~\mathrm{kHz}$ and $\Gamma_{13} = 17.4 \pm 1.2~\mathrm{kHz}$.  The obtained $\Gamma_{13}$ is consistent with the result of $16.1 \pm 0.9~\mathrm{kHz}$ determined from our independent all-optical measurements. The optical decoherence rate obtained from 2PE measurement was $\gamma = 11.0 \pm 1.6~\mathrm{kHz}$, likely including contributions from spectral diffusion induced by the strong control pulses. Substituting these parameters and the measured efficiency into Eq.~(\ref{eq:eff}), and assuming $\eta_{\text{pm}} = 1$ for the co-propagating geometry, gives a control efficiency of  $\eta_{\text{control}} = 82.8\%$. 

To realize backward retrieval, we adopted the pulse sequence shown in Fig.~\ref{fig:fig3}(a). The propagation direction of the $\pi$ pulse at $t_5$ was reversed to satisfy the phase-matching condition for backward retrieval in Eq.~(\ref{eqn:pm}). In addition, the two $\pi$ pulses at $t_5$ and $t_6$ were swapped, and the sequence timing was adjusted so that the background 4LEs were separated from SE through a combination of spatial and temporal discrimination. The corresponding noise pathways are analyzed in detail in the next subsection. The measured backward efficiency is $0.6 \pm 0.01\%$ at $23~\mu\text{s}$. Fitting the echo-decay curves in Fig.~\ref{fig:fig3}(b) yields $\Gamma_{35} = 24.3 \pm 1.2~\text{kHz}$ and $\Gamma_{13} = 16.6 \pm 1.1~\text{kHz}$, consistent with the corresponding values for forward retrieval. This agreement indicates that the decoherence dynamics are essentially independent of the retrieval direction. In this configuration, phase matching cannot be assumed to be perfect, meaning that $\eta_{\text{pm}}$ does not necessarily reach unity. As a result, $\eta_{\text{control}}$ cannot be independently evaluated, and we can only extract the combined quantity $\eta_{\text{pm}}\eta_{\text{control}}^4 = 13.4\%$.

\subsection{Noise analysis}

\begin{figure}[t]
	\includegraphics[width=1.0\linewidth]{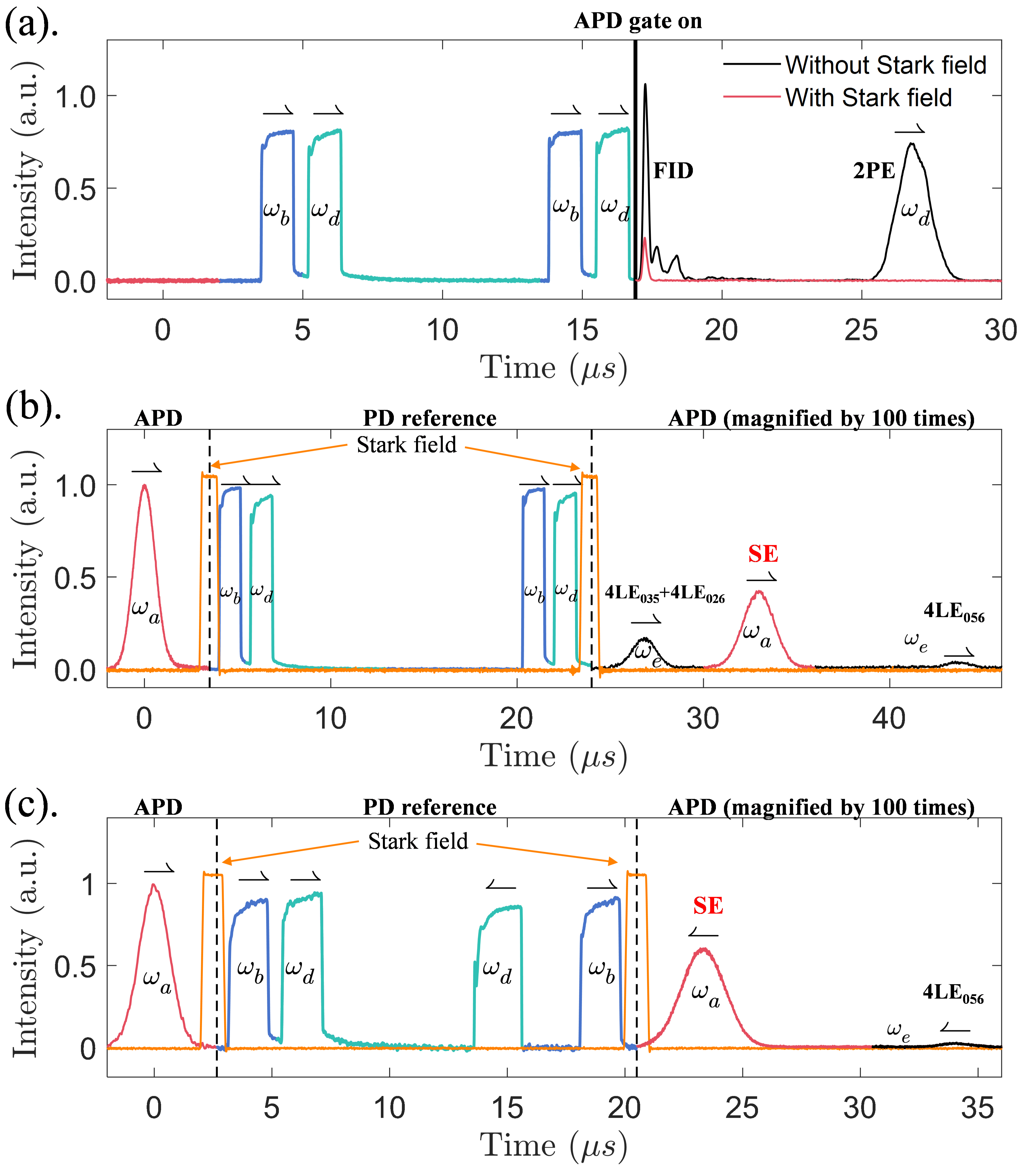}
	\caption{\label{fig:noise} Experimental noise analysis. (a). FID and 2PE noise produced without input signal. (b) and (c) describe the measured 4LE noise in forward and backward schemes, respectively. }
\end{figure}

In the absence of an input signal, both 2PEs and free-induction-decay (FID) noise originate solely from the applied $\pi$ pulses. The 2PEs arise from pulse imperfections associated with the $\omega_b$--$\omega_b$ and $\omega_d$--$\omega_d$ pulse pairs. However, because the population on the $\omega_b$ transition is largely depleted during the state-preparation stage, the corresponding echo contribution is negligible. Applying an electric-field (Stark) pulse at $t_7$ strongly suppresses these coherent noise components. As shown in Fig.~\ref{fig:noise}(a), the pulse nearly eliminates the 2PE noise and reduces the FID noise induced by the final $\pi$ pulse to below 20\% of its original level.

In practice, the control $\pi$ pulses cannot achieve unit transfer efficiency, i.e., $\eta_{\text{\tiny control}}<1$. As a result, the coherence created by the input signal is not completely transferred, and the residual coherence can be further driven by subsequent control pulses, giving rise to additional background. Apart from the intended 4LE generated at $t_4$, the input signal excitation can combine with any $\omega_b$--$\omega_d$ or $\omega_d$--$\omega_b$ $\pi$-pulse pair to form parasitic 4LE sequences. These unwanted 4LEs act as coherent noise and have optical frequencies different from those of the SE. Their emission directions satisfy the phase-matching condition~\cite{beavan2011photon,Sup}
\begin{equation}\label{eqn:4le}
	\mathbf{k}_{\text{4LE}} = -\mathbf{k}_0 + \mathbf{k}_{\pi_1} + \mathbf{k}_{\pi_2},
\end{equation}
where $\mathbf{k}_{\pi_1}$ and $\mathbf{k}_{\pi_2}$ denote the wave vectors of the two $\pi$ pulses involved in generating the corresponding 4LE. We denote a 4LE generated by the input signal at $t_0$ and the $\pi$ pulses at $t_m$ and $t_n$ as $\mathrm{4LE}_{0mn}$; for example, $\mathrm{4LE}_{023}$ is generated by the input signal at $t_0$ and the $\pi$ pulses at $t_2$ and $t_3$.

In the forward-retrieval sequence shown in Fig.~\ref{fig:noise}(b), three corresponding 4LE noise components, $\mathrm{4LE}_{026}$, $\mathrm{4LE}_{035}$, and $\mathrm{4LE}_{056}$, can be generated and detected. According to the 4LE timing condition in Eq.~(\ref{eqn:4le}), the SE and these noise components can be temporally separated by appropriately adjusting the pulse sequence. The first 4LE noise peak in Fig.~\ref{fig:noise}(b) is formed by the overlap of $\mathrm{4LE}_{026}$ and $\mathrm{4LE}_{035}$. In contrast, the backward-retrieval configuration generates two backward-propagating 4LEs and one forward-propagating 4LE. The experimental results are shown in Fig.~\ref{fig:noise}(c), where $\mathrm{4LE}_{025}$ is not observed because it is emitted before the application of the Stark pulse. More details of the noise analysis are presented in Supplemental Materials~\cite{Sup}.

The total transfer efficiency of the SE involves all four $\pi$ pulses, so its intensity scales as
\begin{equation}\label{eqn:SE_intensity}
	\text{I}_{\text{\tiny SE}} \propto \eta_{\text{\tiny control}}^4.
\end{equation}
By contrast, each 4LE noise component undergoes only two $\pi$ pulses. Its intensity therefore scales as
\begin{equation}\label{eqn:4LE_intensity}
	\text{I}_{\text{\tiny 4LE}} \propto (1-\eta_{\text{\tiny control}})^2 \eta_{\text{\tiny control}}^2.
\end{equation}
Details of the derivation are provided in the Supplemental Materials~\cite{Sup}. Therefore, from the measured SE-to-4LE intensity ratio $\text{I}_{\text{\tiny SE}}/\text{I}_{\text{\tiny 4LE}}$ together with the corresponding decoherence factors, we infer a control-pulse efficiency of $\eta_{\text{\tiny control}} = 81.5\%$ for forward retrieval, in agreement with the value obtained from the preceding analysis. For backward retrieval, the same method yields $\eta_{\text{\tiny control}} = 85.3\%$. Using this inferred value, together with the experimentally determined linewidths and decoherence rate in Eq.~(\ref{eq:eff}), we estimate the phase-matching efficiency to be $\eta_{\text{\tiny pm}} \approx 7.1\%$ for the backward-retrieval experiment. Increasing $\eta_{\text{\tiny control}}$ can effectively suppress the 4LE noise. In addition, a filter crystal can be used to eliminate the residual noise~\cite{zhu2024integrated, beavan2012demonstration}, enabling noiseless quantum storage.

\subsection{Qubit fidelity}

\begin{figure}[t]
	\includegraphics[width=1.0\linewidth]{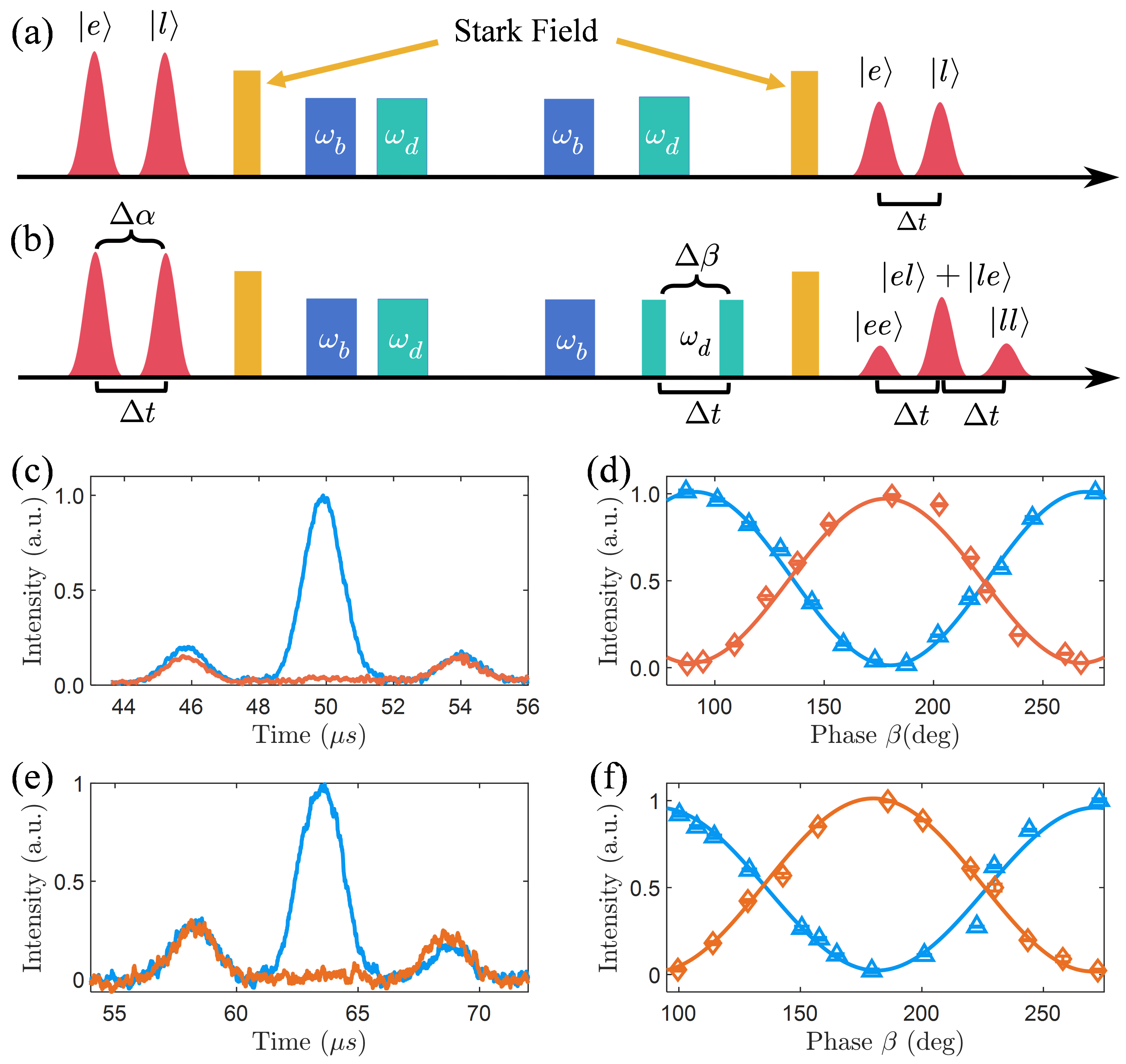}
	\caption{\label{fig:fidelity} Qubit fidelity measurement. (a) illustrates the storage and retrieval process of the $\ket{e}$ and $\ket{l}$ time-bin qubit. (b) shows the interference measurement sequence used to evaluate the coherence between these two qubit states. (c,d) and (e,f) correspond to the fidelity measurements of time-bin qubits for forward and backward retrieval, respectively. The time-bin delays are $\Delta t = 4~\mu\mathrm{s}$ in (c,d) and $\Delta t = 5~\mu\mathrm{s}$ in (e,f). Panels (c) and (e) show the constructive and destructive interference fringes, while panels (d) and (f) show the corresponding interference curves for $\Delta \alpha = 0^\circ$ (blue) and $\Delta \alpha = 90^\circ$ (orange).}
\end{figure}

\begin{table}[b]
	\caption{\label{tab:fidelity}
		Fidelities of forward and backward retrieved qubits.}
	\setlength{\tabcolsep}{2pt}
	\begin{ruledtabular}
		\begin{tabular}{@{~~~}ccc@{~~~}}
			\textrm{Fidelity} & \textrm{Forward} & \textrm{Backward} \\
			\colrule
			$F_e$                   & $98.4 \pm 0.2\%$ & $96.3 \pm 0.4\%$ \\
			$F_l$                   & $96.7 \pm 0.3\%$ & $97.1 \pm 0.4\%$ \\
			$F_+$     				& $97.0 \pm 1.2\%$ & $98.2 \pm 1.0\%$ \\
			$F_-$ 					& $98.6 \pm 0.8\%$ & $97.6 \pm 1.3\%$ \\
			$F_T$                   & $97.7 \pm 0.8\%$ & $97.5 \pm 0.9\%$ \\
		\end{tabular}
	\end{ruledtabular}
\end{table}

The input time-bin qubit was prepared in the state $\ket{\psi}=(\ket{e}+e^{i \Delta \alpha}\ket{l})/\sqrt{2}$, where $\ket{e}$ and $\ket{l}$ denote the early and late temporal modes, respectively, and $\Delta \alpha$ is their relative phase. The two time bins were separated by a fixed temporal interval $\Delta t$, as shown in Fig.~\ref{fig:fidelity}(a). The fidelities for the early and late basis states were calculated from the signal-to-noise ratio $F_{e,l}=(S+N)/(S+2N)$, where $S$ is the signal subtracted noise while $N$ is the noise~\cite{gundougan2015solid}. The average fidelity for the computational-basis states is then given by $F_{\text{poles}}=(F_e+F_l)/2$.

For superposition-state measurements, we employed an unbalanced Mach-Zehnder interferometer (MZI)~\cite{gundougan2015solid,ma2021elimination}. As shown in Fig.~\ref{fig:fidelity}(b), the $\pi$ pulse at $t_6$ was replaced by two $\pi/2$ pulses separated by a delay $\Delta t$ and with a relative phase $\Delta \beta$, enabling interference between the two time bins. From the measured interference fringes, we extracted the visibility $V$ and determined the corresponding superposition-state fidelity as $F_{\mathrm{equator}}=(1+V)/2$. This procedure was performed for input states with $\Delta\alpha=0^\circ$ and $\Delta\alpha=90^\circ$, yielding two superposition-state fidelities. The corresponding interference results for forward and backward retrieval are shown in Fig.~\ref{fig:fidelity}(c,d) and Fig.~\ref{fig:fidelity}(e,f), respectively.

The overall conditional fidelity per qubit was obtained by combining the contributions from the basis-state and superposition-state measurements~\cite{gundougan2015solid}:
\begin{equation}
	F_T=\frac{1}{3}F_{\text{poles}}+\frac{2}{3}F_{\text{equator}}.
\end{equation}
All results are summarized in Table~\ref{tab:fidelity}, yielding an overall conditional fidelity of $97.7 \pm 0.8\%$ for forward retrieval and $97.5 \pm 0.9\%$ for backward retrieval.

\subsection{Discussion}

\begin{figure}[t]
	\includegraphics[width=\linewidth]{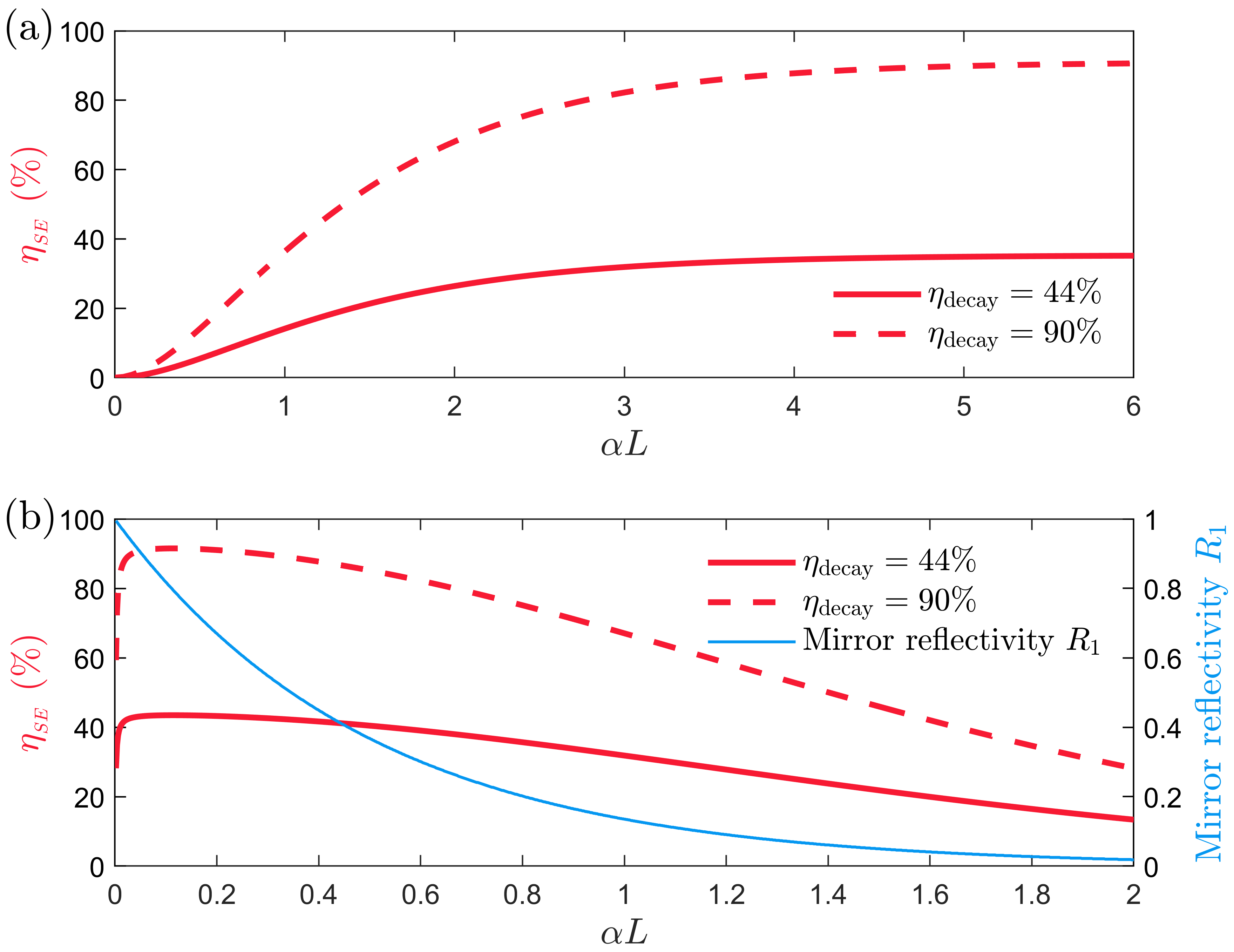}
	\caption{\label{fig:fig4} Theoretical memory efficiency estimation of (a). backward-retrieval and (b). cavity-enhancement, with different $\eta_{\text {decay}}$ and under the assumption of $\eta_{\text {\tiny control}}=1$ and $\eta_{\text{\tiny pm}}=1$. The blue curve in (b) represents the maximum achievable efficiency under different optical depths, while the blue curve denotes the first mirror reflectivity $R_1$ required to achieve this maximum efficiency. $R_2$ is set to a typical value of 0.999~\cite{afzelius2010impedance}.  }
\end{figure}

The limited backward-retrieval efficiency can be primarily attributed to two experimental constraints. First, imperfect alignment and spatial mode overlap between the counter-propagating beams limit the phase-matching efficiency $\eta_{\text{\tiny pm}}$. Second, the available control-field power does not fully support complete population transfer, thereby reducing the overall control efficiency $\eta_{\text{\tiny control}}$. Both limitations can be addressed using integrated photonic structures. Waveguide confinement provides superior spatial mode overlap and significantly boosts the control-field intensities, enabling shorter control pulses and efficient adiabatic population-transfer techniques~\cite{rippe2005experimental_CHS, roos2004quantum_CHS, tian2011reconfiguration_HSH}. This approach can realistically push both $\eta_{\text{\tiny pm}}$ and $\eta_{\mathrm{control}}$ toward unity while further suppressing residual 4LE noise~\cite{Sup}.

Fig.~\ref{fig:fig4}(a) shows the expected backward-retrieval efficiency $\eta_{\text{\tiny SE}}$, calculated using the experimentally determined $\eta_{\text{\tiny decay}} \approx 44\%$ at a storage time of 23~$\mu$s. Achieving high-efficiency backward retrieval primarily requires improving both $\eta_{\text{\tiny retrieval}}$ and $\eta_{\text{\tiny decay}}$~\cite{zhu2024integrated,liu2025millisecond}. Because backward emission eliminates reabsorption, $\eta_{\text{\tiny retrieval}}$ can be enhanced directly by increasing the optical depth. Recent advances in isotopically purified $^{151}\mathrm{Eu}^{3+}:\mathrm{Y}_{2}\mathrm{SiO}_{5}$ indicate that substantially higher optical depths can be achieved without sacrificing coherence~\cite{zhu2024integrated,liu2025millisecond}. These developments suggest that stoichiometric rare-earth crystals may be a promising option for future implementations of the present scheme, owing to their potentially favorable combination of high optical depth and narrow linewidths~\cite{ahlefeldt2013minimizing,vasilenko2025optically}.

Cavity enhancement provides a complementary route to improve the memory efficiency. We consider an optical cavity formed by an input coupling mirror and an end mirror, with reflectivities \(R_1\) and \(R_2\), respectively. Under impedance-matching conditions, the transmission of the input coupling mirror matches the absorption of the intracavity medium, such that the incident field is efficiently coupled into the cavity and interacts repeatedly with the ensemble~\cite{afzelius2010impedance, sabooni2013efficient}. Owing to the destructive interference between the promptly reflected field and the intracavity leakage field, nearly all of the incident light can be absorbed by the sample, thereby effectively enhancing the light-matter interaction. Neglecting spectral dispersion and inhomogeneous broadening~\cite{moiseev2010efficient,moiseev2021broadband}, Eq.~(\ref{eq:eff}) becomes
\begin{equation}\label{eqn:cavity}
	\eta_{\text{\tiny retrieval}}=
	\frac{4d^2 e^{-2d}(1-R_1)^2 R_2}
	{\left(1-\sqrt{R_1R_2}e^{-d}\right)^4},
\end{equation}
As shown in Fig.~\ref{fig:fig4}(b), cavity enhancement enables near-unity efficiency even at moderate optical depths, making it particularly suitable for low-doping systems in which \(\eta_{\mathrm{decay}}\) is already high and the optical depth remains the primary bottleneck.

\section{Conclusion}

In summary, we report the first experimental demonstration of backward retrieval in a spin-wave quantum memory while retaining the full optical depth of the ensemble and suppressing coherent noise. SE modulation enables deterministic phase control without population removal, thereby providing geometrical flexibility and full optical-depth utilization that are not readily available in AFC- or NLPE-based protocols. These results demonstrate the feasibility of high-fidelity backward retrieval and suggest a practical route toward higher efficiency through improved backward retrieval and its combination with cavity enhancement. Overall, our work identifies SE protocol as a promising platform for efficient backward retrieval and highlights its potential for scalable, long-lived quantum-network nodes based on rare-earth-ion materials.

\begin{acknowledgments}
This work was supported by the Quantum Science and Technology-National Science and Technology Major Project (QNMP, Grant No. 2021ZD0301204), the National Natural Science Foundation of China (Grant Nos. 11904159, 12004168 and 12304454), National Key Research and Development Program of China (Grant No. 2022YFB3605800), Guangdong Innovative and Entrepreneurial Research Team Program (Grant No. 2019ZT08X324), Guangdong Basic and Applied Basic Research Foundation (Grant No. 2021A1515110191), the Key-Area Research and Development Program of Guangdong Province (Grant No. 2018B030326001), Department of Science and Technology of Guangdong Province (No. 2020B0303050001) and The Science, Technology and Innovation Commission of Shenzhen Municipality (KQTD202108110900-49034).
Z. X. and M. G. contributed equally to this work.
\end{acknowledgments}

\bibliography{reference}

\end{document}